# Prime Focus Spectrograph - Subaru's future -

Hajime Sugai*[a], Hiroshi Karoji[a], Naruhisa Takato[b], Naoyuki Tamura[a], Atsushi Shimono[a], Youichi Ohyama[c], Akitoshi Ueda[d], Hung-Hsu Ling[c], Marcio Vital de Arruda[e], Robert H. Barkhouser[f], Charles L. Bennett[f], Steve Bickerton[g], David F. Braun[h], Robin J. Bruno[h], Michael A. Carr[g], João Batista de Carvalho Oliveira[e], Yin-Chang Chang[c], Hsin-Yo Chen[c], Richard G. Dekany[i], Tania Pereira Dominici[e], Richard S. Ellis[j], Charles D. Fisher[h], James E. Gunn[g], Timothy M. Heckman[f], Paul T. P. Ho[c], Yen-Shan Hu[c], Marc Jaquet[k], Jennifer Karr[c], Masahiko Kimura[c], Olivier Le Fèvre[k], David Le Mignant[k], Craig Loomis[g], Robert H. Lupton[g], Fabrice Madec[k], Lucas Souza Marrara[e], Laurent Martin[k], Hitoshi Murayama[a], Antonio Cesar de Oliveira[e], Claudia Mendes de Oliveira[l], Ligia Souza de Oliveira[e], Joe D. Orndorff[f], Rodrigo de Paiva Vilaça[e], Vanessa Bawden de Paula Macanhan[e], Eric Prieto[k], Jesulino Bispo dos Santos[e], Michael D. Seiffert[h,j], Stephen A. Smee[f], Roger M. Smith[i], Laerte Sodré Jr[l], David N. Spergel[g], Christian Surace[k], Sebastien Vives[k], Shiang-Yu Wang[c], Chi-Hung Yan[c]

[a]Kavli Institute for the Physics and Mathematics of the Universe, The University of Tokyo, 5-1-5, Kashiwanoha, Kashiwa, 277-8583, Japan;
[b]Subaru Telescope, National Astronomical Observatory of Japan, 650 North A`ohoku Pl., Hilo, Hawaii, 96720, USA;
[c]Institute of Astronomy and Astrophysics, Academia Sinica, P.O. Box 23-141, Taipei, Taiwan;
[d]National Astronomical Observatory of Japan, 2-21-1 Osawa, Mitaka, Tokyo 181-8588, Japan;
[e]Laboratório Nacional de Astrofisica, MCTI, Rua Estados Unidos, 154, Bairro das Nações, Itajubá, MG, Brazil;
[f]Department of Physics and Astronomy, Johns Hopkins University, 3400 North Charles Street, Baltimore, MD 21218, USA;
[g]Department of Astrophysical Sciences, Princeton University, Princeton, NJ, 08544, USA;
[h]Jet Propulsion Laboratory, 4800 Oak Grove Drive, Pasadena, CA 91109, USA;
[i]Caltech Optical Observatories, 1201 East California Blvd., Pasadena, CA 91125 USA;
[j]Astronomy Department, California Institute of Technology, 1200 East California Blvd, Pasadena, CA 91125, USA;
[k]Observatoire Astronomique de Marseille-Provence, Laboratoire d'Astrophysique de Marseille, Pôle de l'Etoile Site de Château-Gombert 38, rue Frédéric Joliot-Curie 13388 Marseille cedex 13, France;
[l]Instituto de Astronomia, Geofisica e Ciencias Atmosfericas, Universidade de São Paulo, Rua do Matão, 1226 - Cidade Universitária - 05508-090, Brazil;

## ABSTRACT

The Prime Focus Spectrograph (PFS) of the Subaru Measurement of Images and Redshifts (SuMIRe) project has been endorsed by Japanese community as one of the main future instruments of the Subaru 8.2-meter telescope at Mauna Kea, Hawaii. This optical/near-infrared multi-fiber spectrograph targets cosmology with galaxy surveys, Galactic archaeology, and studies of galaxy/AGN evolution.

Taking advantage of Subaru's wide field of view, which is further extended with the recently completed Wide Field Corrector, PFS will enable us to carry out multi-fiber spectroscopy of 2400 targets within 1.3 degree diameter. A microlens is attached at each fiber entrance for F-ratio transformation into a larger one so that difficulties of spectrograph

* hajime.sugai@ipmu.jp; phone 81 4 7136-6551; fax 81 4 7136-6576; http://member.ipmu.jp/hajime.sugai/

Updated 1 March 2012

design are eased. Fibers are accurately placed onto target positions by positioners, each of which consists of two stages of piezo-electric rotary motors, through iterations by using back-illuminated fiber position measurements with a wide-field metrology camera. Fibers then carry light to a set of four identical fast-Schmidt spectrographs with three color arms each: the wavelength ranges from 0.38 μm to 1.3 μm will be simultaneously observed with an average resolving power of 3000.

Before and during the era of extremely large telescopes, PFS will provide the unique capability of obtaining spectra of 2400 cosmological/astrophysical targets simultaneously with an 8-10 meter class telescope. The PFS collaboration, led by IPMU, consists of USP/LNA in Brazil, Caltech/JPL, Princeton, & JHU in USA, LAM in France, ASIAA in Taiwan, and NAOJ/Subaru.

## 1. BACKGROUND

Prime Focus Spectrograph (PFS) is an optical/near-infrared multi-fiber spectrograph with 2400 fibers, each of which is set onto a target position quickly by a fiber positioner using two-staged rotational motors. It is planned to be mounted on the Subaru 8.2-meter telescope at Mauna Kea, Hawaii. The fibers patrol within 1.3 degree diameter, a significant portion of the extended field of view with the recently completed Wide Field Corrector (WFC). PFS will share this WFC with Hyper Suprime Cam (HSC), a counterpart with the imaging capability in the Subaru Measurement of Images and Redshifts (SuMIRe) project.

In 2011 January the PFS has been endorsed by Japanese community as one of the main future instruments of the Subaru telescope. This decision was made based on PFS Science White Paper, which included spectrograph design and perspectives with PFS on cosmology with galaxy surveys, Galactic archaeology, studies of galaxy evolution and high redshift galaxies, and AGN studies including its evolution and dust-enshrouded population, and also based on detailed scientific/technical discussions in Subaru users meetings. Furthermore, in 2011 December, a Memorandum of Understanding between The National Astronomical Observatory of Japan (NAOJ) and The Institute for the Physics and Mathematics of the Universe (IPMU) on the PFS Project has been signed. In this agreement, the NAOJ supports the PFS program lead by the IPMU as an international project in its design, construction and observation. Before and during the TMT era, this optical/near-infrared multi-fiber spectrograph will provide the unique capability of obtaining spectra of 2400 cosmological/astrophysical targets simultaneously with an 8-10 meter class telescope.

The PFS collaboration to pursue this relatively large project consists of experienced collaborators: Caltech/JPL in the USA, who have been intensively developing a new type of fiber positioner, Princeton & JHU in the USA, who have many experiences of building imagers and spectrographs including SDSS and FUSE, LAM in France, who has also been experienced with large spectrographs such as MUSE, USP/LNA in Brazil, one of whose specially strong fields are fibers, ASIAA in Taiwan, who has been involved tightly in SuMIRe/HSC project in its filter exchanger and optical testing system, NAOJ/Subaru, who owns and operates the Subaru telescope as well as many instruments mounted on it, and IPMU, who has led this project with Hitoshi Murayama as the PI. In order to organize this international collaboration efficiently, PFS project office has been established since 2011 April.

Science goals of PFS have been further discussed and extended by Japanese community and in science working groups, which formed in 2011 August. There are presently four working groups, including cosmology, Galactic archaeology, galaxy evolution, and AGNs. Following the Subaru advisory committee's decision made in 2011 November on Subaru strategic program (SSP), we are now considering submitting a proposal of these programs together as a single SSP in order to maximize the output impacts achieved with this unique instrument.

The specification of PFS has been determined through active interactions between technical and science teams. Table 1 shows current optical/near-infrared multi-fiber spectrographs for a comparison. A bright F-ratio of 2.2 from the WFC is even more challenging compared with the brightest F-ratio of 2.5 among optical spectrographs. A microlens is attached at each fiber entrance for F-ratio transformation into a larger one, 2.8, so that difficulties of spectrograph design are eased. In FMOS, a near-infrared multi-fiber spectrograph on Subaru, a bright F-ratio of 2 is transformed into 5 in large fiber connectors[1]. Fibers are accurately placed onto target positions by two-staged piezo-electric rotary motors through iterations by a wide-field metrology camera. Fibers then carry light to a set of four identical fast-Schmidt spectrographs with three color arms each: the wavelength ranges from 0.38 μm to 1.3 μm will be simultaneously observed with an average resolving power of 3000. Table 1 shows that our camera F-ratio of 1.1 is also challenging: it is the smallest (brightest) F-ratio. We are planning to use double Schmidt corrector for ensuring good image quality for this bright camera. We are considering possibilities of also having a medium resolution mode of resolving power of 5000 or larger for red arms. Table 2 summarizes the basic characteristics for PFS.

In 2012 March we had the Conceptual Design Review of the PFS project. The review panel, consisting of five experienced external reviewers, recommended that the project proceed to the next phase. Science part of the Conceptual Design Review has been revised and been recently published in astro-ph[2]. In this conference we have seven papers on the PFS instrumentation: instrument overview (the present paper), fiber system[3], fiber positioner[4], spectrograph[5], dewar & detector[6], metrology camera[7], and system software[8]. In the present paper we describe the instrument overview. Details of individual components are found in the corresponding component-dedicated papers. We are scheduling the PFS technical first light on the Subaru as in 2017 and the start of SSP survey as in 2018.

Table 1. Comparison with current optical/near-infrared multi-fiber spectrographs. The F-ratio of the telescope of 2.2 and the F-ratio of camera of 1.1 for the PFS are one of the smallest (brightest). We will challenge these small F-ratios with uses of a microlens for F-ratio transformation and of double Schmidt corrector for a camera.

| Telescope | Instr. | Prime Focus? | F-ratios/ Pixel size | Number of fibers | Fiber diameter/ Positioner |
|---|---|---|---|---|---|
| Subaru 8.2m | PFS [0.38-1.3μm] | YES | 2.2 → 2.8 with microlens (collimator 2.5; camera 1.1)/ 15 μm | 2400 (1.3 deg FOV) | 128 μm: 1".1/ Two rotary motors |
| VLT 8.2m | FLAMES/ GIRAfEE | no | 15 → 5 with microlens (camera 1.2)/ 15 μm | 600 | 230 μm: 1".2/ Magnetic button |
| MMT 6.5m | Hectospec | no | 5 (camera 1.5)/ 13.5 μm | 300 | 250 μm: 1".5/ 5-axis robot |
| WHT 4.2m | AF2 | YES | 2.5 13.5 μm | 150 | 90 μm: 1".6/ Robot Autofib2 |
| Guoshoujing 4m | LAMOST | YES | 5 (collimator 4; camera 1.2)/ | 4000 (5 deg FOV) | 320 μm: 3".3/ Two rotary motors |
| AAT 3.9m | 2dF | YES | 3.5 (collimator 3.15; camera1.2)/ 24 μm | 400 | 140 μm: 2".1 |
| WYIN 3.5m | Hydra | no | 6.3 (collimator 5; camera 1.8) 12 μm | 100 (288 slots) | 200 μm: 2"/ Magnetic button |
| SDSS III 2.5m | APOGEE | no | 5 (collimator 3.5; camera 1.4) | 300 | 120 μm: 2" |
| SDSS III 2.5m | BOSS | no | 5 | 1000 (3 deg FOV) | 120 μm: 2" |
| AAO UKST 1.2m | 6dF | YES | 2.5 | 150 | 100 μm: 6".7/ Off-telescope robot |
| Subaru 8.2m | FMOS [NIR: 0.9-1.8μm] | YES | 2 → 5 in fiber connectors (collimator 4.7; camera 1.5)/ 18 μm | 400 (0.5 deg FOV) | 100 μm: 1".2/ Echidna spine |

Table 2. Basic characteristics for PFS.

| Basic characteristics for PFS | |
|---|---|
| **Field element** | |
|   Shape | Plane; thickness 52 mm |
| **Fiber** | |
|   Number | 2400, including fiducial fibers |
|   Diameter | Core 128 μm; Cladding 170 μm; Buffer 190 μm |
| |     Core size corresponds to 1".13 at field center and 1".03 at corner when a microlens attached |
|   Connectors | Two positions: at telescope spider & near spectrographs |
| **Microlens** | |
|   Shape | Plano-concave (aspheric); thickness 600 μm; glued to fiber input edge |
|   F-ratio transformation | F/2.2 to F/2.8 |
| **Fiber positioner** | |
|   Positioning mechanism | Two stages of rotary motors |
|   Positioner distribution | Hexagonal pattern |
|   Distance from neighboring positioners | 8.0 mm |
|   Patrol region | 9.5 mm diameter circle for each fiber (fill factor 100% with overlaps) |
| **Field shape & size** | |
|   Field shape | Hexagon |
|   Field size | Diagonal line length |
| |     i) Hexagon defined by fiber positioner centers (i.e., twice of distance from field center to farthest fiber positioner centers): 448.00 mm = 1.366 deg on sky |
| |     ii) Hexagonal patrol region within which any astronomical target can be accessed at least with one fiber: 453.92 mm = 1.383 deg on sky |
| | Effective diameter of circle whose area is equal to |
| |     i) hexagon defined based on fiber positioner centers: 407.41 mm = 1.248 deg on sky |
| |     ii) hexagon defined based on patrol region: 412.83 mm = 1.264 deg on sky |
| **Spectrograph** | |
|   Number | 4 spectrographs, each with a slit of 600 fibers & 3-color arms: located on tertiary mirror floor infrared side |
|   Slit length | ~140 mm, with center-to-center fiber spacing of 230 μm |
|   F ratios | All-Schmidt type: collimator F/2.5 & camera F/1.1 |
|   Grating | VPH; diameter 280 mm |
|   Wavelength region | 380-1300 nm (blue: 380-670 nm; red: 650-1000 nm; NIR 970-1300 nm) |
|   Spectral resolution | ~2.7 A |
| **Dewar & Detector** | |
|   Dewar window | Camera Schmidt corrector |
|   Pixel size | 15 μm |
|   Detector | A pair of 2K x 4K fully depleted CCDs for each of blue & red arms; 4K x 4K HgCdTe (1.75 μm cutoff) for NIR arm |
| **Metroloy camera** | |
|   Location | At Cassegrain |
|   Magnification | ~0.04 |
|   Camera aperture size | ~130 mm; diffraction-limited image quality |
|   Detector | 120M 2.2μm-pixel CMOS sensor |

## 2. INSTRUMENT OVERVIEW

Figure 1 shows a schematic view of PFS. 2400 fibers and their corresponding fiber positioners are mounted on the prime focus of the Subaru telescope. Light entering fibers is fed into four spectrographs located on the tertiary mirror floor (infrared side). The instrument consists of several components, as shown in PFS block diagram (Figure 2). Individual components are described in the following subsections.

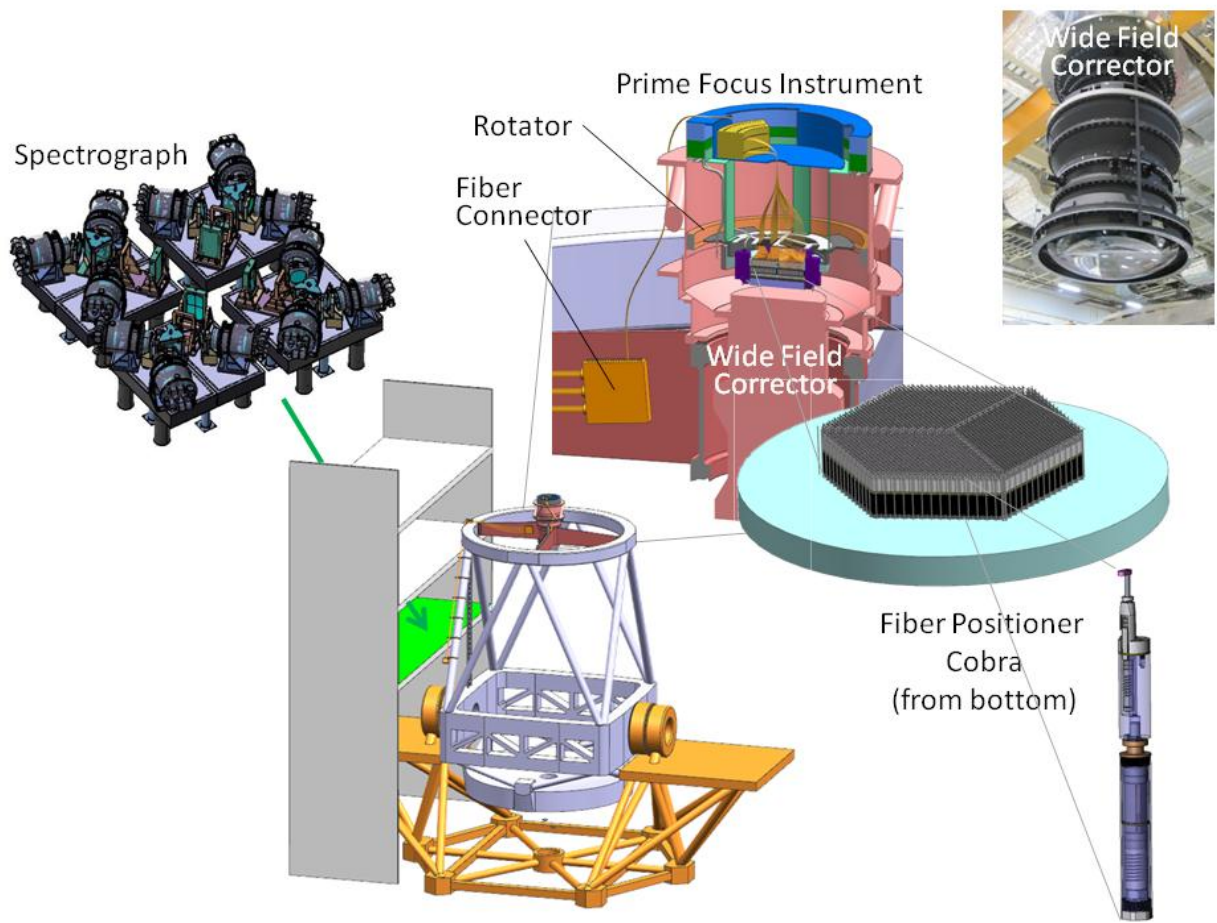

Figure 1. PFS is an optical/near-infrared multi-fiber spectrograph with 2400 fibers and their corresponding fiber positioners that are mounted at the prime focus of the Subaru telescope, feeding four spectrographs mounted on the tertiary mirror floor (infrared side). The metrology camera will be placed at Cassegrain.

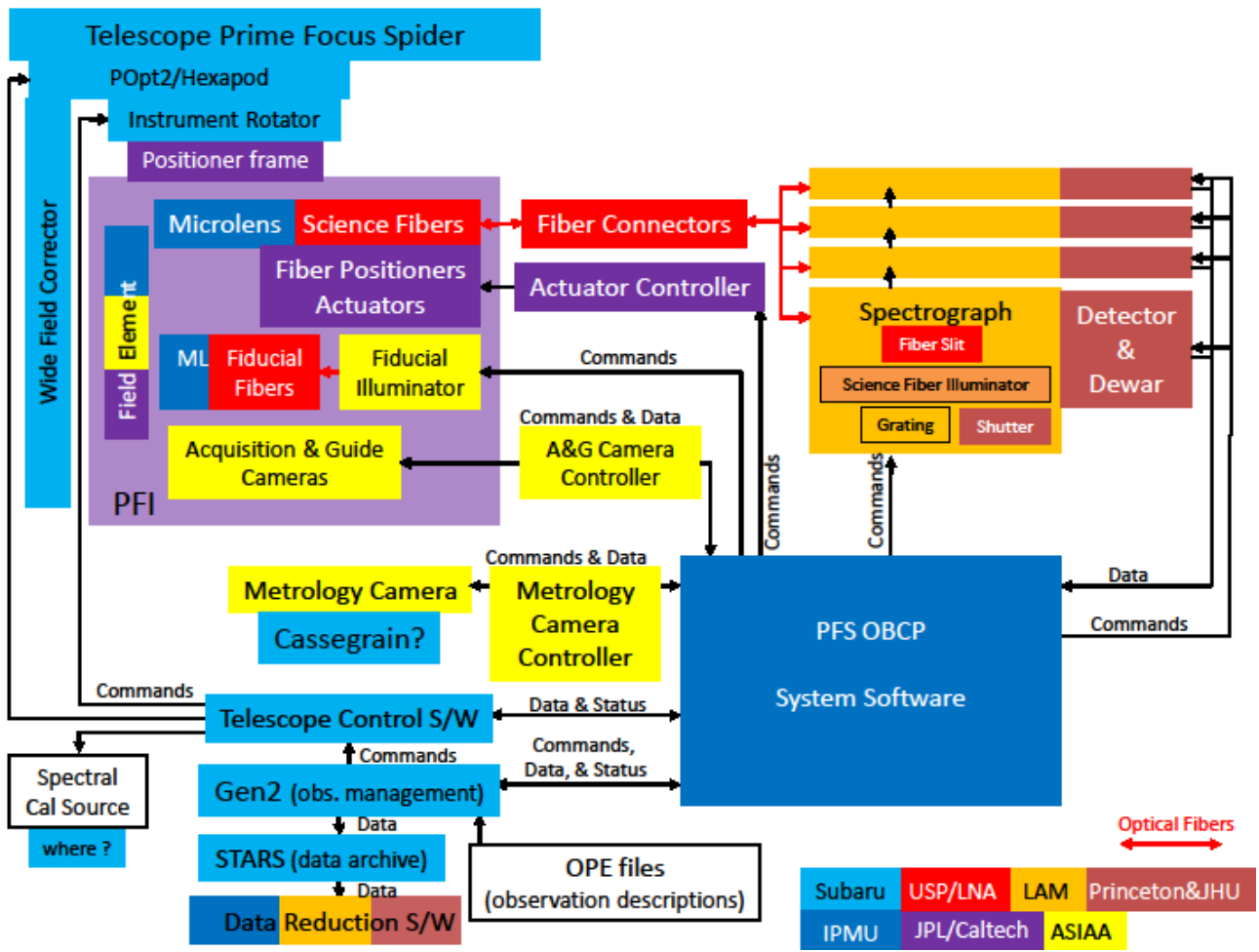

Figure 2. PFS block diagram showing the required components as well as their responsible collaboration institutions.

**2.1 Sharing Wide Field Corrector with Hyper Suprime Cam and using Field Element**

PFS shares several elements with HSC. These include Prime Focus Unit (PFU) called POpt2, field rotator, hexapod, and Wide Field Corrector (WFC). Since the HSC filter and dewar window are not used in PFS, the total thickness of these elements is compensated with PFS Field Element, a flat glass of 52 mm thickness. This field element is located close to the telescope prime focus and makes a correction for spherical aberration, which otherwise would be introduced by the removal of the filter and dewar window.

**2.2 Fiber system & Fiber positioner**

After going through WFC and PFS Field Element and just before the prime focus, light reaches a microlens attached to each fiber entrance edge. The plano-concave microlens transforms the F-ratio of 2.2 to a fainter (or slower) one of 2.8, which ensures an efficient acceptance of light and also eases the spectrograph design (Figure 3). We are presently considering a possibility of producing these microlenses with the molding technique of glasses. Aspherical surfaces are easily produced in the molding technique, which provides more flexibility on the microlens design compared with the polishing technique. The molding also gives the uniformity of the microlens qualities. This is essential to the uniform sensitivity among fiber channels. The molding glasses gives wider selection of available refractive indices compared with the molding plastics. It also guarantees longer life time.

Each fiber tip position is controllable in-plane by a piezo-electric Fiber Positioner, nicknamed "Cobra." Each fiber tip can be positioned within its 9.5 mm diameter circular patrol region (Figure 4). The patrol regions are in a hexagonal close-packed pattern, with 8 mm separation, and fill a hexagonally shaped 1.3 degree field of view. The overlap between adjacent patrol regions enables 100% sky coverage of this hexagonal field. The hexagonal shape of the field of view allows efficient tiling of the sky for large area surveys. The fibers can be completely reconfigured for a new field in 40 seconds, with each fiber tip placed to an accuracy of 5 microns (corresponding to ~0.054 arcsecond on the sky) although the actual reconfiguration time may depend on the exposure time for the metrology camera (section 2.5) required for obtaining stable back-illuminated fiber images against the dome seeing. Fiber tips are translated (rather than tilted) within the image plane, providing uniform coupling efficiency for all configurations. Some fraction of fibers which are not allocated to astronomical targets will measure the spectrum of the sky. Intensive experiments and evaluation have been carried out on the prototype Cobras[3].

We are planning to use 128, 170, 190 μm, respectively, for the fiber core, cladding, and buffer diameters. We have intensively measured characteristics of fibers with the similar diameters and diameter ratios, including absolute transmittance and F-ratio degradation (Figure 5). Through a microlens, light produces an astronomical target image onto the input edge of a fiber and is relayed with a 55-meter length fiber to one of four spectrographs placed on the tertiary mirror floor (infrared side). The total number of 2400 fibers is divided into four groups, each of which provides a slit of 600 fibers on the corresponding spectrograph. Fiber Connectors provide flexibility in instrument exchange, as well as testing, and retain the possibility of using the fiber system to feed other instruments. The fiber connectors are used in two places: at a side of a telescope spider and near the spectrographs. On the telescope side, the connectors allow most of the fiber bundle to be permanently installed on the telescope while facilitating the removal of the POpt2 as necessary. At the spectrograph end, the connectors provide easier and safer installation of spectrographs as well as a way to test the instrument with local sources. They also are important in that a possible independent higher resolution spectrograph can be realized relatively easily, making use of the Cobra and the fixed fiber system.

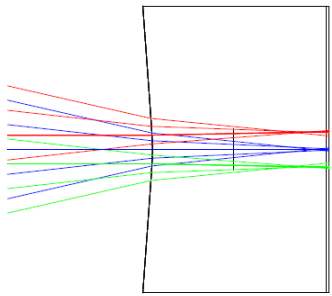

Figure 3. A plano-concave microlens attached to each fiber entrance transforms the bright input F-ratio of 2.2 (left side of this figure) into a fainter one of 2.8. This ensures the efficient light acceptance and eases the spectrograph design.

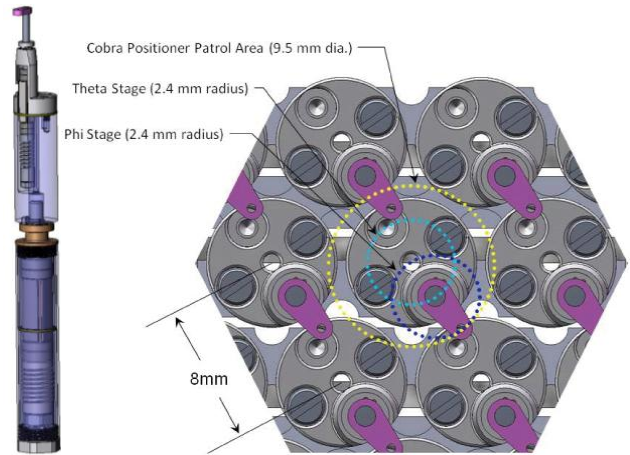

Figure 4. Fiber positioner "Cobra" consisting of two-stage piezo-electric rotary motors. This two-motor rotation provides a total of 9.5 mm-diameter patrol region. Fiber positioners are distributed in a hexagonal shape with 8-mm separations from six neighboring positioners each. This combination of the configuration of positioners and the patrol region of each positioner gives 100% sky coverage of the whole hexagonal field, with some fraction overlapped by two positioner patrol regions and a tiny fraction overlapped by three positioner patrol regions.

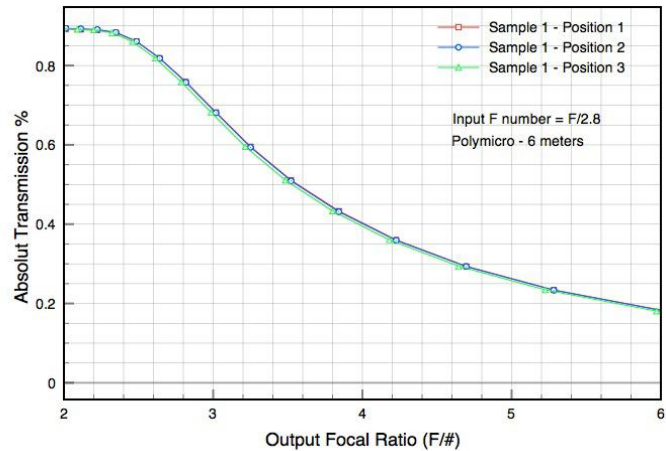

Figure 5. An example of absolute transmittance and F-ratio degradation measurements. The absolute transmission integrated within a given output F-ratio cone is shown for the input light with F-ratio of 2.8. Measurements for this 6-meter fiber sample were carried out three times for the purpose of stability test, with the fiber moved after each measurement.

## 2.3 Spectrograph

Light relayed through a fiber reaches one of four identical spectrographs (Figure 6). In each spectrograph, 600 fibers are linearly aligned in a slit assembly and output beams are collimated by a reflector with the F-ratio of 2.5. This is slightly brighter than the transformed F-ratio of 2.8 by a microlens in order to avoid light loss due to a possible F-ratio degradation by a fiber. After divided into three wavelength arms by two dichroic mirrors, beams are corrected by a Schmidt collimator corrector in each arm. Each beam is dispersed by a VPH grating and its spectra are imaged with a reflector camera with the F-ratio of 1.1 after corrected by Schmidt camera correctors, one of which works also as a dewar window as described in section 2.4. The magnification of 0.44 (=1.1/2.5) should produce a 56 μm fiber image, whose original core size is 128 μm, without aberration. This corresponds to 3.75 pixel on a 15 μm-pixel detector. The specification of spectrograph design is as the followings and this is satisfied almost on all area of detector in our design.

The Ensquared Energy (EE) for a fiber image is:
≥50% within a square of 3 x 3 pixels for each spectral band;
≥90% within a square of 5 x 5 pixels for each spectral band.

With four identical spectrographs, each of which consists of three arms, we simultaneously obtain 2400 spectra covering a wide wavelength region ranging from 0.38 μm to 1.3 μm. While it is cooled down to the detector operating temperatures inside the dewars as described in section 2.4, the temperature of the spectrograph part outside the dewars, including fiber slit, collimator, dichroics, and gratings, will be controlled around 0 degree Celsius in order to reduce the thermal background contributions particularly in NIR. Our one-year monitoring results of temperature and humidity at the tertiary mirror floor infrared side and optical side as well as inside the HDS room on the optical side suggest that we will achieve this level of cooling with the use of air conditioner and some enclosure/room structure.

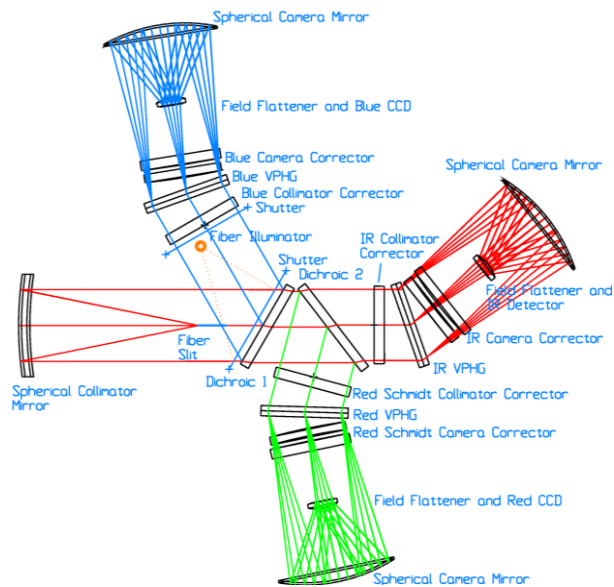
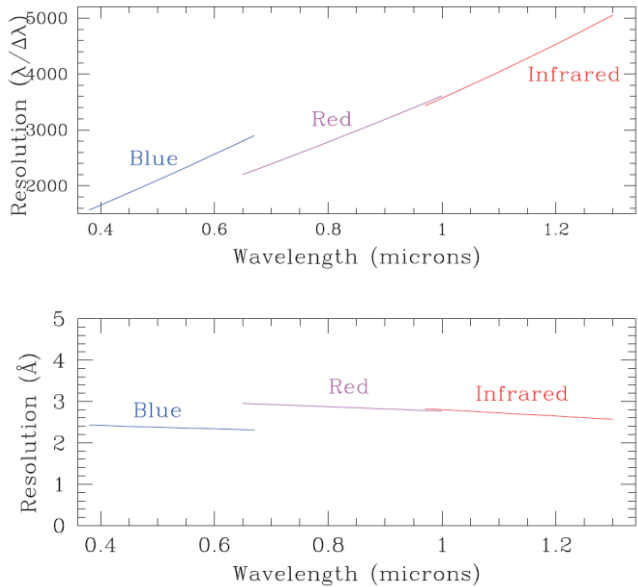

Figure 6. (left) Schematic diagram of one of four identical spectrographs. Light output from a fiber in the fiber slit, which is a 600-fiber array actually vertical to this figure plane, is collimated by a spherical reflector. Two dichroics divide light into three wavelength arms. (right) Designed resolution (expressed as a function in the upper panel, and Angstroms in the lower panel) for each of the three arms of the spectrograph.

## 2.4 Detector & Dewar

In order to reduce schedule risks and cost, we use the same dewar structure and the similar associated cooling system for optical and NIR arms (Figure 7). A Schmidt corrector lens for camera is used also as a dewar window to avoid an additional throughput loss by a normal plate dewar window. In each of blue and red arms, a pair of edge-buttable 2K x 4K fully depletion region CCDs is used for assuring the high sensitivity particularly in the red wavelength region. Anti-reflection coatings are optimized for the corresponding wavelength region for each of blue- and red-arm CCDs. The 600 spectra are dispersed along the 4K-pixel columns, so the gap falls between spectra. For each NIR arm, a single 4K x 4K pixel near-infrared detector is used. The cutoff wavelength is set as 1.75 μm, as short as possible in order to reduce the thermal background. We also use thermal background cutoff filters to suppress thermal contributions down to 1.3 μm. The tilt/focus mechanism for a detector uses three levers in either a manual or an electric way: the movement of each of these levers is transformed into one hundredth movement of the corresponding point of the detector.

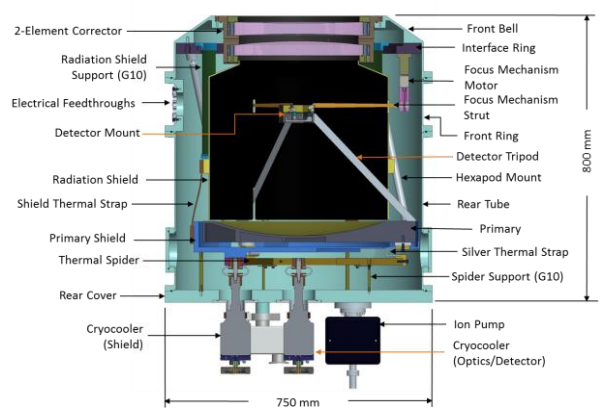

Figure 7. One of double Schmidt corrector for a camera is used also as a dewar window. Focus mechanism struts do not make an additional obscuration since they are hidden within the projected areas already obscured by tripods for supporting the detector.

**2.5 Measurement of accurate fiber position and its feedback to fiber positioner**

A metrology camera mounted in a Cassegrain container (or in the Cassegrain slit-viewer layer) looks up at back-illuminated fiber tips through WFC (Figure 8). This determines the location of the fiber tips, allowing iterations of positioning of the fibers on the selected science targets. The metrology camera consists of imaging optics with a CMOS detector and includes an internal flat field source. Calibration of geometrical distortion inherent in the optics of the metrology camera is achieved by having illuminated fixed fiducial fibers whose positions are known accurately. The fiducial fibers also allow correction for geometric distortion in the WFC, through measurement and modeling. The field of view of the metrology camera is designed to image all the fibers in one exposure. We are now estimating the optimized exposure time, investigating the characteristics of the Subaru dome seeing by using back-illuminated FMOS fibers. The centroid of each fiber image is calculated and used to determine each science fiber position with respect to the fixed fiber locations. The science fibers are back-illuminated by sources located inside the four spectrographs.

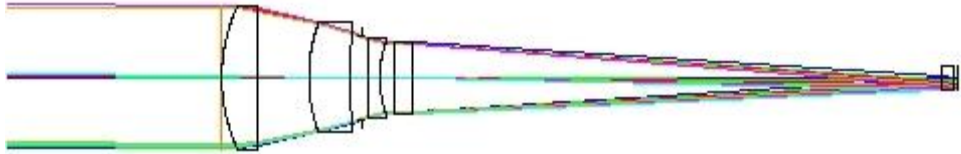

Figure 8. Optical layout of the metrology camera, which is planned to be mounted in Cassegrain container. The total length is around 645 mm.

**2.6 Software and execution processes of observations**

The observation preparation software parses large surveys into a series of individual observations, matches fibers to designated targets, selects appropriate guide stars, predicts observing times, and provides sequence files in a Subaru-format. The PFS OBCP controls the entire instrument under the Subaru observation management software Gen2. Execution of an observation begins with moving the telescope to the desired field center. While the telescope is slewing, the fiber positioning system executes commands from the OBCP to simultaneously move all fibers to their required positions, with the real positions verified by the metrology system. With several iterations, the position error of each science fiber can be effectively minimized. In parallel with this process, the acquisition & guide system refines the pointing of the telescope and angle of instrument rotator. The hexapod is commanded to the appropriate position to maintain the WFC alignment in the presence of flexure in the POpt2 structure. Once positioning is complete, the autoguiding of the telescope starts. The on-site data reduction system produces quick-look information on the completed observation so that data quality and survey progress can be monitored. The collected raw data will be transferred to and archived in the STARS data archive system, which then provides data retrieval for further scientific data reduction and analysis.

# 3. EXPECTED PERFORMANCE

**3.1 Optical parameter design flow and fiber core aperture size determination**

Although the detailed optical design of the instrument is accomplished through iterations among many parameters considered, rough values as the starting point of the most important parameters are rather directly derived from scientific motivations and technical constraints as the followings:

(1) Fiber core diameter ~ 100 μm
This is derived from the image scale at the prime focus provided through WFC, 10"-11" per mm, and the typical target galaxy size. Small adjustments for this core size are possible with the use of a microlens at the fiber input edge.

(2) Numerical aperture of fiber ~ 0.22
This is derived from the F-ratio of the beam provided from WFC, 2.2. Small adjustments are possible with the use of a microlens.

(3) Camera F-ratio ~ 1.1

This is derived based on the detector pixel size of 15 μm for both optical and NIR arms. The de-magnified fiber core image size is far oversampled. In order to avoid this oversampling as much as possible, it is required to have a large de-magnification factor and our challenge is to design a small Camera F-ratio of 1.1 for this purpose.

(4) Use of microlens

The above discussion is too simplified: we actually need to take into considerations non-telecentricity and vignetting by WFC and the F-ratio degradation of a fiber. In order to make the above parameters work realistically, we use a microlens to transform the input F-ratio from 2.2 to 2.8.

Figure 9 (left) shows the estimated signal-to-noise (S/N) ratio variation for the typical target galaxies with the *i*-band (around 0.76 μm) AB magnitude of 22-23, as a function of fiber-core aperture radius. For this purpose, the HST/ACS images on the GOODS-N region (http://archive.stsci.edu/prepds/goods/) was analyzed and the median growth curve of these galaxies was used. As the sky brightness, 19.7 AB magnitude arcsec$^{-2}$ in the *i* band was taken from the Subaru/Suprime-Cam web site. A larger fiber core aperture includes a more signal from a target galaxy while the background noise increases linearly with a root square of aperture. The optimized fiber-core aperture on the sky has been determined as about 1".1 diameter for these typical target galaxies with $i_{AB}$ = 22-23 mag in the typical seeing conditions around 0".6 or so at full width of half maximum (FWHM). This fiber core diameter corresponds to 100 μm at the prime focus through WFC and to 128 μm when a microlens is attached at the fiber input edge. Figure 9 (right) shows that this core aperture diameter of 1".1 is optimized in worse seeing conditions for stars, such as 0".8.

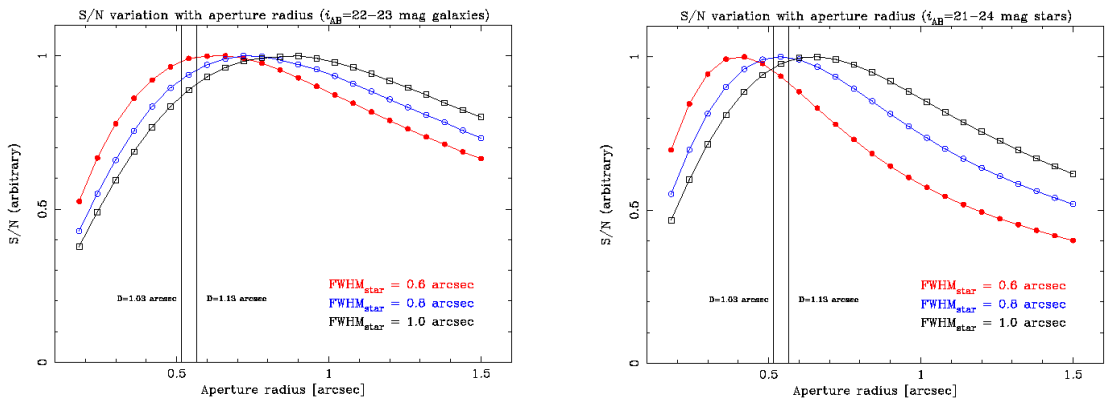

Figure 9. (left) Normalized signal-to-noise ratios for $i_{AB}$ = 22-23 magnitude galaxies as a function of fiber-core aperture radius. The sky brightness of $i_{AB}$ = 19.7 mag arcsec$^{-2}$ was used. Vertical lines show the PFS fiber-core radius, which corresponds to 1".13 diameter at the field center and 1".03 at the field corner. The seeing sizes of 0".6, 0".8, and 1".0 were assumed respectively for filled circles, open circles, and open squares. (right) Normalized signal-to-noise ratios for $i_{AB}$ = 21-24 magnitude stars as a function of fiber-core aperture radius.

## 3.2 Total throughput

Figure 10 shows the assumed total throughput including the atmosphere to detector, which has been used to estimate the S/N ratio variation in section 3.1. Figure 10 (left) shows the best case. i.e., at the zenith and for the field-center fiber while Figure 7 (right) shows the worst case, i.e., at elevation of 30 degree and for the field-corner fiber. We will have around ~10-20% of total throughput depending on the wavelength and observing conditions, with even lower cases in the shortest and longest wavelengths.

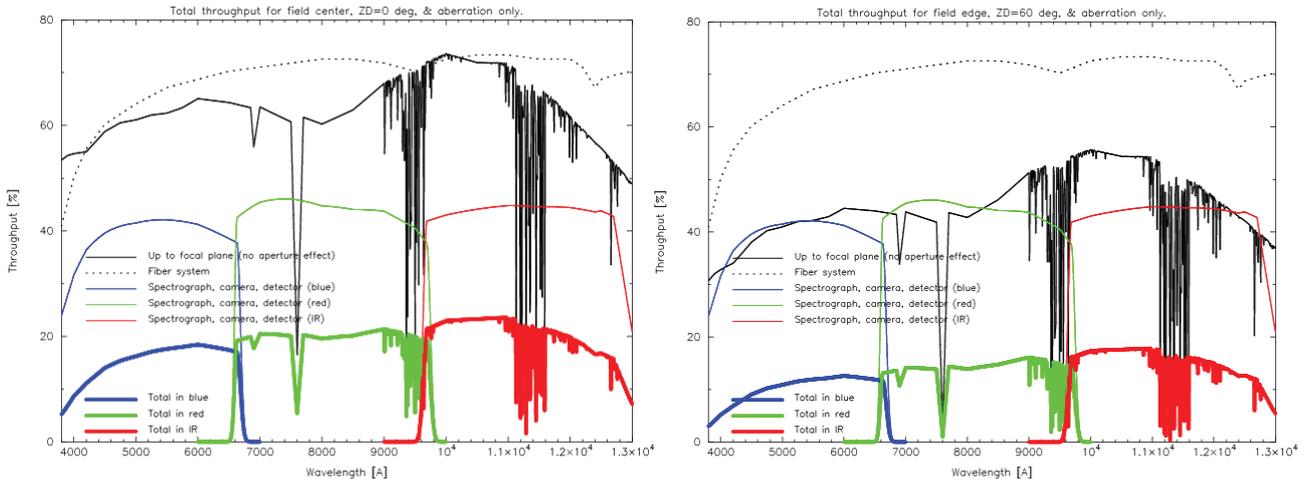

Figure 10. (left) Predicted total throughput for the zenith including the atmosphere to detector, including its quantum efficiency. (right) Predicted total throughput for the telescope elevation of 30 degree. This also includes the vignetting by the WFC.

### 3.3 Signal-to-noise ratio

Figure 11 shows the predicted S/N ratios for a point source with an AB magnitude of 22.5, including the effect of losses due to the fiber-core aperture. This aperture effect depends on wavelength due to small residual chromatic aberrations of the WFC Atmospheric Dispersion Corrector and on the telescope elevation, and on the assumed seeing size. Moffat profile with an FWHM of $0.64(\lambda/\lambda_0)^{-0.2}$, where $\lambda_0 = 0.73$ μm and the exponent parameter β of 4.765, was assumed as the seeing. Fiber positions were set on the 0.76 μm image. Relatively uniform S/N ratios are obtained over the wide observed wavelength range. This is essential to the efficient observations with the balanced depths.

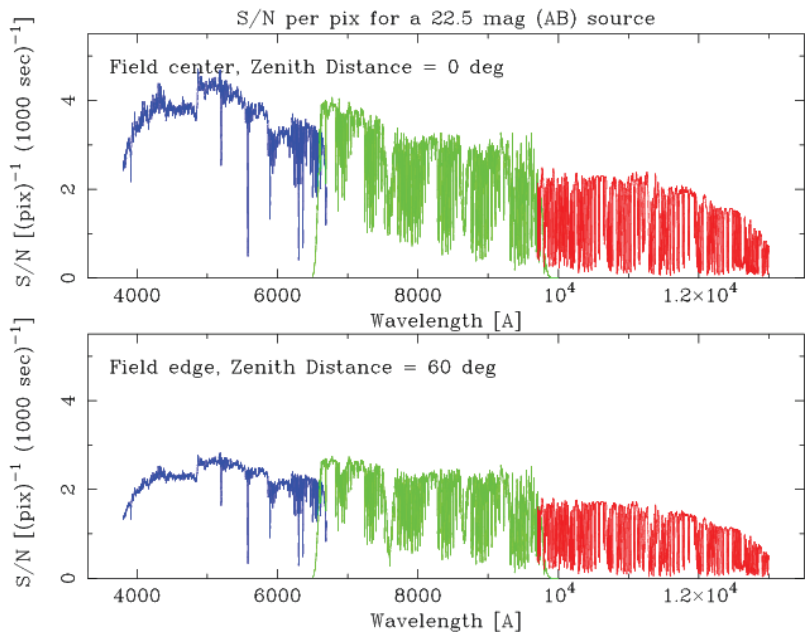

Figure 11. Predicted S/N ratios for a point source with an AB magnitude of 22.5 for the PFS as a function of wavelength. Moffat profile with an FWHM of $0.64(\lambda/\lambda_0)^{-0.2}$, where $\lambda_0 = 0.73$ μm and an exponent parameter β of 4.765, was assumed as the seeing. A single exposure with the integration time of 1000 sec and the sky-subtraction by using ten sky fibers were assumed. The upper curve is the best case, i.e., for a fiber at the field center and with the telescope pointing at the zenith, while the lower curve is the worst case, i.e., for a fiber located at the field corner and with the telescope elevation of 30 degree.

## 4. CONCLUSIONS

The Prime Focus Spectrograph (PFS), an optical/near-infrared multi-fiber spectrograph targeting cosmology with galaxy surveys, Galactic archaeology, and studies of galaxy/AGN evolution, is one of the main future instruments for the Subaru telescope. It enables us to simultaneously obtain 2400 targets' spectra within 1.3 degree diameter with 100% sky coverage and to cover the wavelength range between 0.38 and 1.3 μm. The PFS is a challenging instrument in terms of bright F-ratios of input light from the telescope and of the Schmidt camera, and also in the accurate and quick fiber positioning by using a new-type robotic fiber positioners and a wide field metrology camera. Our design including a microlens and double Schmidt corrector, actual experiments on the proto-type fiber positioners, and measurements of fiber properties and the Subaru dome seeing characteristics show that it is feasible to accomplish the required instrument performance.

## ACKNOWLEDGEMENTS

We gratefully acknowledge support from the Funding Program for World-Leading Innovative R&D on Science and Technology (FIRST) program "Subaru Measurements of Images and Redshifts (SuMIRe)", CSTP, Japan." We appreciate staff members at Subaru Telescope for continuously supporting our activities. We also acknowledge the WFMOS-B team whose accumulated efforts of many years have inspired us.